\documentclass[lettersize,journal]{IEEEtran}
\usepackage{amsmath,amsfonts}
\usepackage{algorithmic}
\usepackage{algorithm}
\usepackage{array}
\usepackage[caption=false,font=normalsize,labelfont=sf,textfont=sf]{subfig}
\usepackage{textcomp}
\usepackage{stfloats}

\usepackage{url}
\usepackage{verbatim}
\usepackage{graphicx}
\usepackage{cite}
\begin{document}

\title{A Smart Contract based Crowdfunding Mechanism for Hierarchical Federated Learning}

\author{{Hongze~Liu$^{1}$,
        Jie~Li\IEEEauthorrefmark{1}$^{,2}$,~\IEEEmembership{Senior Member,~IEEE,}
        Shijing~Yuan$^{2}$, Wenqi~Cao$^{1}$,
        and~Bowen Li$^{2}$}\\ \IEEEauthorblockA{$^1$  UM-SJTU JI,  Shanghai Jiao Tong University, Shanghai, China}\\
        \IEEEauthorblockA{$^2$  Department of Computer Science and Engineering,  Shanghai Jiao Tong University, Shanghai, China}\\
        \IEEEauthorblockA{seniordriver233@sjtu.edu.cn, lijiecs@sjtu.edu.cn, 2019ysj@sjtu.edu.cn,
        davidcwq@sjtu.edu.cn,
        li-bowen@sjtu.edu.cn },
\thanks{* Jie Li is the corresponding author.}
\thanks{This work has been submitted to the IEEE for possible publication. Copyright may be transferred without notice, after which this version may no longer be accessible.}}



\maketitle

\begin{abstract}
Hierarchical Federated Learning (HFL) is introduced as a promising technique that allows model owners to fully exploit computational resources and bandwidth resources to train the global model. However, due to the high training cost, a single model owner may not be able to deploy HFL.
To address this issue, we develop a smart contract based trust crowdfunding mechanism for HFL, which enables multiple model owners to obtain a crowdfunding model with high social utility for multiple crowdfunding participants. 
To ensure the authenticity of the crowdfunding mechanism, we implemented the Vickey-Clark-Croves (VCG) mechanism to encourage all crowdfunding participants and clients to provide realistic bids and offers.
At the same time, in order to ensure guaranteed trustworthiness of crowdfunding and automatic distribution of funds, we develop and implement a smart contract to record the crowdfunding process and training results in the blockchain. We prove that the proposed scheme satisfies the budget balance and participant constraint.
Finally, we implement a prototype of this smart contract on an Ethereoum private chain and evaluate the proposed VCG mechanism. The experimental results demonstrate that the proposed scheme can effectively improve social utility while ensuring the authenticity and trustworthiness of the crowdfunding process.
\end{abstract}

\begin{IEEEkeywords}
 Federated learning, VCG Mechanism, Crowdfunding, Truthfulness, Smart contract.
\end{IEEEkeywords}

\section{Introduction}
Currently, Federated Learning (FL) has been widely applied in industry as a promising distributed learning solution \cite{FL21}. However, the bandwidth bottleneck between the client and the central parameter server limits the development of FL \cite{TWC20HFL1}. To overcome this shortcoming, hierarchical federation learning (HFL) is proposed, aiming to alleviate the bandwidth constraint by reducing the number of communications between the client and the central parameter server \cite{TWC20HFL2}. A typical hierarchical federation learning consists of three entities, including model owners, aggregators, and clients. But a single model owner may not be be able to implement the training for a particular model due to lack of resources or profit, e.g., model owners who are limited with capability cannot deploy a global model independently \cite{TWC21HFL}.

\begin{figure*}[!t]
\centering
\includegraphics[width=6.5in]{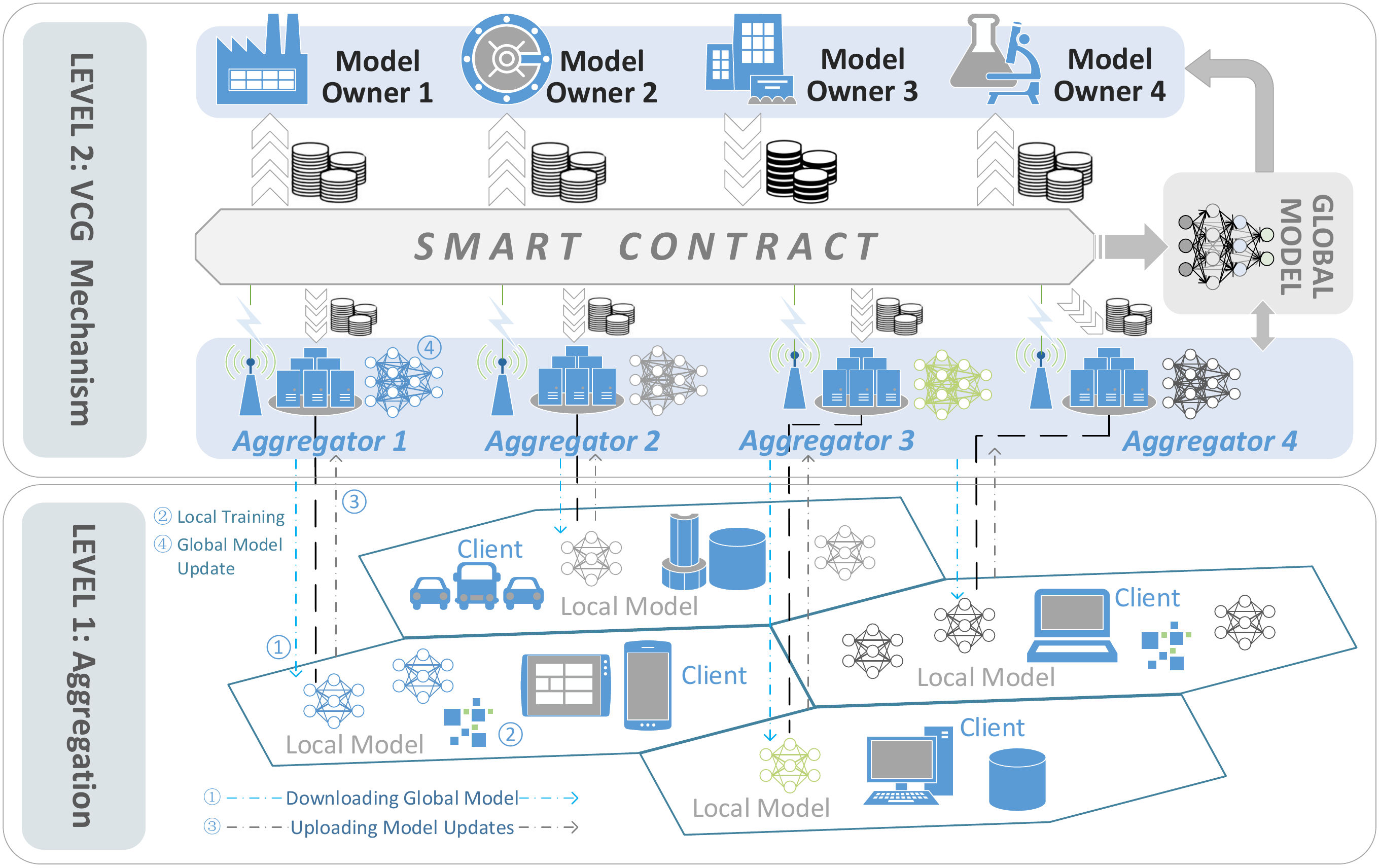}
\caption{The System Framework of HCFL.}
\label{fig_1}

\end{figure*}

\indent
To bridge this gap, an intuitive approach to assist these model owners is to establish a mechanism of model crowdfunding, which enable multiple model owners to cooperate and train a shared global model in HFL by aggregating resources and apportioning cost. 
Developing a model crowdfunding mechanism for multiple model owners faces two main problems. The first problem stems from \emph{the fairness of the model bidder and the tenderer}.
Furthermore, due to the difference in demand of model among all model owners, the second problem is \emph{how to determine a model that can maximize the overall utility}. Therefore, a mechanism that can obtain the true evaluation from participants, both owners and clients, is needed in order to assess the value of a particular model by comparing the margin between the individual expected utility and cost.\\ 
\indent
The fairness problem of the model crowdfunding has been discussed in previous worsk. Lyu \emph{et al.,} \cite{Lyu2020} propose a framework of FPPDL to achieve the fairness by sharing gradient among the model owners. However, this method assumes the model is pre-determined. To our best known, this is the first paper discussing the selection of model in FL.\\
\indent
On the other hand, as far as the truthfulness is concerned, many existed works mainly focus on the honesty of the clients by implementing contract\cite{Ding2021}, auction\cite{Jiao2021} and game theory\cite{Lim2022}. However, these approaches will  cause monopoly and dishonesty between the model owner and the client due to the unilateral exposure of the information of the client(e.g. the cost, data quality). Consequently, the model owner can exploit all margin utility from FL training by only allocating the payment that can just compensate the cost instead of their true utility. Therefore, the foundation of the truthfulness is based on the equal status of information that no participants have extra knowledge of others' evaluation of the model,  capability of training and so forth.\\
\indent
In this article, we propose a smart contract based crowdfunding FL architecture to address the above two problems. As Fig. \ref{fig_1} shown, the target model will be selected from model owners model proposals based on other participants' assessment to it. The architecture leverage the advantage of the Block chain which protects the privacy and neutrally operate with resources based on the delicate smart contract. The smart contract will blind the privacy among the participants. The smart contract compares the margin utility among all the proposed models according to the claim of value from model owners and the claim of the cost from clients. Furthermore, as a neutral operator, the smart contract can enforce the payment and other procedure during the training. In terms of truthfulness, the application of VCG mechanism enable the system to select the model that can obtain the maximum social utility in this common goods auction scenario. Considering the complexity of the VCG mechanism, the deployment of Hierarchy Federated Learning structure will not only lightens the communication and non-i.i.d problems which previous works have discussed, but also validates the VCG mechanism by limiting the number of the participants from numerous edge devices to countable base stations which significantly reduces the complexity. By combining the advantage of VCG mechanism, smart contract and HFL, the proposed mechanism can achieve the greatest social well-being in this particular circumstance. \\

\indent
Our contribution are summarized as follows:
\begin{itemize}
    \item[1] We propose a smart contract-based novel trustful mechanism in model crowdfunding FL scenario where multiple model owners and clients participates the common model training.
    \item[2] We design a feasible structure of HFL for the proposed VCG-mechanism which reduce the complexity, preserve privacy and lighten the drawback of VCG mechanism.
    \item[3] We demonstrate the feasibility of this proposed framework in truthfulness and efficiency.
    \item[4]We implement the framework on  CIFAR-10 and verify the encouragement of cooperation of the VCG-HFL architecture.

\end{itemize}

To our best knowledge, this article the first work discussing the truthful crowdfunding mechanism in FL.

\section{Related Work}
Original Federated Learning framework may suffer from the bottleneck of communication inefficiency such as node failures and device dropouts in dynamic process. Bryaan Lim \textit{et al}. proposed a decentralized learning based system model inspired by the Hierarchy Federated Learning with stable two-level resource allocation and incentive design framework in order to lighten  the communication inefficiency  with the help of cluster head[1].\\
\indent
Vickey-Clarke-Croves mechanism is frequently applied in Public Economics as an approach to conduct truthful common goods auction. 
Thi et al.\cite{Thi2021} illustrates the implementation of VCG machanism in FL is a NP-hard problem and  provides a approximation algorithm for it.
Zhang \textit{et al}. also implement a approximation of this mechanism to construct a faithful edge federated learning with the tradeoff among privacy, convergence and payment accuracy loss\cite{Zhang2021}. In this article, the concept of cluster is implemented as the method to approach the VCG mechanism due to the complexity of $K+1$ of this mechanism where $K$ denotes to the number of the participants. Note the idea of cluster in FL is similar to HFL which also clustering the edge devices. Therefore, HFL is a potentially ideal framework to implement the VCG mechanism. However, the VCG mechanism applied in this work emphasizes more on the incentive and faith in training rather than the truthfulness in public auction.\\ 
\indent
Smart contract is crucial in coordination of a group of participants connected through a block chain. Wu \textit{et al}. proposed a Block-chain based trusted collaborative services framework, BlockEdge\cite{Wu2020}. This work leverage the smart contract which enable the BlockEdge to achieve trustworthiness of the collaboration process. Shi  et al.\cite{Shi2018} implements a blockchain smart contract for the FTIM to solve the problem of untruthfulness and difficulty in enforcing payment in MPCS systems. \\
\indent
Most of the work related in FL incentive mechanism mentioned are discussed the framework applying during the training process and assume the model to be trained is determined previously. However, the problem about how to decide the training model fairly among all participants' expectation in crowdfunding scenario is remained to be explored. 

\section{Truthful Mechanism Design}


In this section,  we first introduce the challenges in the crowdfunding scenario in FL and the VCG mechanism which is designed for this circumstance. Then, The details of the framework of the FL architecture powered by a delicate smart contract which is compatible with the proposed VCG mechanism.

\subsection{Crowdfunding challenges}
In this article, we assume all participants are cooperative but selfish, which means their individual utility is prior to the collective interests. This will lead to dishonest and irresponsible behaviour during the crowdfunding process.

\subsubsection{Dishonesty and Unfairness of the Participants}

\indent
 The supply of public goods usually can reach the maximum efficiency due to the difference between the individual margin utility and the overall margin utility. For instance, free-rider problem describes a circumstance where an individual tend to behave sluggishly and benefit from the public goods  if others contribute considerably to this goods.\\
 \indent
 Furthermore, if one side of participants owns an unequal position and forms a monopoly, it can exploit all margin utility which usually will cause inefficiency and unfairness. The unequal positive usually appears when the information(e.g. the knowledge of others' cost, evaluation and so forth), the market power(i.e. the number of competitors) and so forth.
 A traditional FL framework is a typical monopoly example due to the fact that the model owner stands on the unequal positive to the clients. The model owner only need to pay for the cost of the clients rather than the true assess to the model.

\subsubsection{Irresponsible Behaviour}
Without punishment, the selfishness will encourage the participants have irresponsible behaviours since these behaviour usually will reduce the individual cost and increase its utility in public goods decision. In the crowdfunding scenario, the model to be trained is selected from all the proposes from the model owners. Technically, it is hard to evaluate all proposed models with little cost  because of the communication bottleneck, the model assessment technique and the unknown training condition. Therefore, distrust will appear among the model owner if they do not take responsibility to their proposed model which may not be able reach the expected performance. 



\subsection{VCG Mechanism Design}

As previously discussed, dishonesty behaviour will impede the acquire of the maximum social utility. Therefore, To solve the dishonesty problem, a Vickrey-Clarke-Groves(VCG) mechanism is designed which proved to be effective in similar public goods auction\cite{Varian}. \\
In this scenario, the social utility can be defined as the margin between the sum of the value that the training model can yield for each individual and the cost and consumption needed during the training process.

\subsubsection{Clarke Tax}
\indent
VCG mechanism is a group of mechanisms
which can provide incentive for participants to tell the truth in auction. Based on the particular situation we are facing, a feasible approach is to use Clarke Tax from VCG to guarantee the   authenticity of the final result\cite{Clarke}.\\
\indent
In Clarke Tax mechanism, the participants are divided into two groups depending on whether they are the pilot. The pilot is the participant whose participation changes the final outcome (e.g. by claiming a comparatively high or low bid) and, therefore, will be required to pay the tax to compensate the loss of others due to its participants. Meanwhile, everyone including the pilots should pay for the cost which is independent to their bid. Therefore, due to the independence between the bid and the payment and the potential connection between the bid and the tax, truth telling become the weak dominant strategy for the participants.\\
\indent
Some assumption is useful in this particular condition. 
\begin{itemize}
  \item[1.] The participants are risk neutral and rational.
\item[2.] Quasi-linear for the monetary utility.

\end{itemize}{}

\subsubsection{Clarke Tax mechanism design}
\indent
In this particular situation, for each proposed model $m$, the participant $i$ from the group of $I$ owns its estimated true value $v_i^m$, claims the value as $b$(may be equal to $v$ due to dishonesty) and receives a final payment $c$. $I$ can be either the model owner group $O$ or the base station  group $B$.\\ 
\indent
Note that $v,b,c$ can be positive or negative depending on whether it is an owner or a base station. For model owner the $v$ and $b$ is positive as utility and $c$ is negative as the payment to rewarding pool. On the contrary, for base station, the $v$ and $b$ is negative as contribution and $c$ is positive as rewards. \\
\indent
let $m' = \arg\max(|\sum\limits_{I/ i} b_i^m- \sum\limits_{I/ i} c_i^m|)$ represents the most valued model without the influence from the claim of the participant $i$. Note the VCG auction will be held in both model owner side and base station side independently.\\
In this scenario, the pilot is the participant that the final model will be different if it does not join the union. we design the Clarke Tax as following:\\

\begin{table}[]
\centering
\caption{The common notation used in Model owner and Base station.}
\begin{tabular}{l|ll}
\hline
  & Model owner      & Base station    \\ \hline
c & Cost(-)          & Reward(+)       \\
b & Claimed value(+) & Claimed cost(-) \\
v & True value(+)    & True cost(-)    \\ 
$C_m$ & Total cost for m & Total reward for m\\
p & Capability    & Contribution\\
I(Group notation) &  O          &  B\\
\hline

\end{tabular}

\label{tab:my-table}
\end{table}

\begin{itemize}
    \item [1.] if $i$ is not the pilot, which indicates $m = m'$, it should pay $c_i^m$  if the model is chosen otherwise the payment is 0.
    \item[2.] if there is a model eventually being chosen(e.g. the claim of the value excesses the claim of the cost) and $i$ is the  pilot, it should pay a tax of $|(\sum\limits_{I / i} b_j^{m} + \sum\limits_{I / i} c_j^{m})- (\sum\limits_{I / i} b_j^{m'}+ \sum\limits_{I / i} c_j^{m'})|$, where $j \in I/i$ indicates the group of $I$ except $i$.
    \item[3.]if finally all models' value can not afford the cost and $i$ is the pilot, it should pay a tax of $|\sum\limits_{I / i}c_j^m + \sum\limits_{I / i}b_j^m|$.
\end{itemize}{}

 we will use $Tax(b_i,b_{-i},c)$ to denote the enforced tax payment according to this rule in the latter part of the article.\\
 \indent
 The payment $c$, determined by the smart contract and being independent to the claim, can be assigned in  many ways depending on the belief of fairness or incentive (e.g. equally allocation $c_i=\frac{C_m}{N}$  or allocating by capability or contribution $c_i = \frac{C_m p_i}{\sum p}$ ). \\
 \indent
 Although VCG mechanism can guarantee the truthfulness there are also some obstacle of this mechanism such as the complexity, the budget balance problem and so forth. Therefore, to implement this mechanism in FL, a compatible system framework is needed which will be introduced in the next section. \\

\subsection{Hierarchy Chained Federated Learning }

To implement the VCG mechanism and solve the irresponsible problem, we design a Hierarchy Chained Federated Learning(HCFL) which can enforce the payment by a smart contract with punishment mechanism.

\subsubsection{The HCFL Architecture}
Hierarchy Federated Learning(HFL) is proposed to solve the communication inefficiency existing in the normal FL. Instead of cooperate with numerous edge devices. HFL establishes countable base stations which are aimed to cluster edge devices. The updates from edge devices are first aggregated to their corresponding base station and ,then, the base stations process the obtained data and upload the local model to the model owner for final aggregation.  \\
\indent
In the scenario of crowdfunding, the model owners are also not unique. Therefore,  a block chain with a delicately designed smart contract can collaborate the model owners and the base stations. The block chain can record all behaviour from the participants and implement punishment based on that when facing malicious action.

\subsubsection{Smart Contract}



\indent
The smart contract can enforce the payment and execute the training procedures. Therefore, to implement the VCG mechanism a delicate smart contract is needed. Particularly, the missions for the smart contract are 
\begin{itemize}
    \item[1] Blind direct information from each  participant.
    \item[2] Announce necessary and privacy-free information to either side. 
    \item [3]Collect the Clarke Tax and recycle it. 
    \item[4]Reward and Punish the participant according to their behaviour.
    \item[5]Gather the payment  from owners and allocate it to clients.
\end{itemize}{}

\begin{figure}[!t]

\includegraphics[width=3.5in]{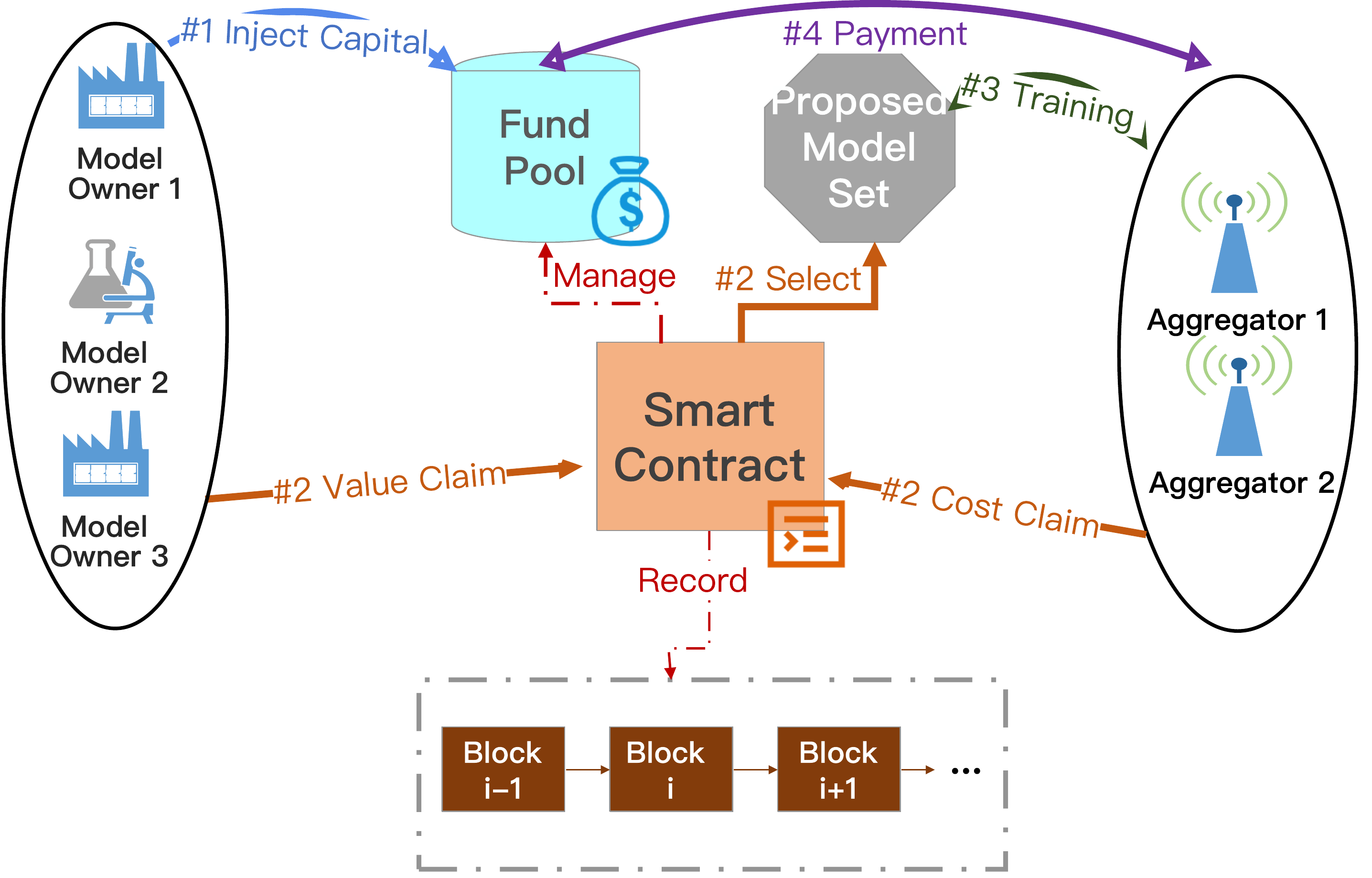}
\caption{The Smart Contract in HCFL}
\label{fig_2}
\end{figure}

The design of the smart contract is shown in Fig.\ref{fig_2} and the working flow for the whole procedure of HCFL is detailed below.

\subsubsection{Working Flowing}
All participants should inject capital to smart contract before the training start. \\
\indent
First, Model owners propose their ideal models' and their descriptions $m(\omega,M,acc,t,y)$ which should contain the trainable parameters $\omega$, the model characteristics $M$, the expected accuracy $acc$, the training round $t$ and the target set $y$. Via the smart contract, other model owners will obtain the claim of $acc$ and $y$ to determine their utility $\{v^i_m(acc,y), i \in O\}$ and the base station will get the parameters of $\omega$,$t$,$M$ to estimate the cost $\{v^B_m(\omega,t,M), i \in B\}$. Note that the $v$ are determined by participants based on individual rationality whose criterion may vary among them.\\
\indent
Then, the smart contract collect the claim of cost $\{b^i_m,i \in B\}$ from BS and the bid of price $\{b^i_m,i \in O\}$ from MO for each proposed model. Based on VCG mechanism, the smart contract selects the model which owns the largest margin between the sum of the cost and the bid. \\
\indent
After the selection of the model, the smart contract announce the winner from the model owners whose proposed model is selected eventually. The particular model owner deliver the model to the base station and start the HFL training. \\
\indent
At the end of the training, the winner evaluate the contribution from each base station and share the global model to other model owners. The rest of model owners will determine whether to punish the winner based on the performance of the shared model(e.g. Actual accuracy) and report their assessment to the smart contract which will be recorded in the block.\\
\indent
Then, the smart contract selects a total amount of payment $C_m$ randomly but with budget balance(e.g. within the interval of the sum of the cost and the bid $\sum\limits_{i \in O} b_i^m \geq C_m \geq \sum\limits_{i \in B} b_i^m$), determines the tax, punishment and reward proportion based on those information and executes all payments $\{c^i_m,i \in \{O,B\}\}$ from the down payment from all participants.

\section{Analysis of the system performance  }

\subsection{Truthfulness and Efficiency}

\subsubsection{Truth telling in auction}
Clarke has provided the prove for the Clarke Tax mechanism enable telling the truth to become the weak dominant strategy\cite{Clarke}. \\
\indent let $U_i(b_i,b_{-i})$ denotes the eventual utility function for participant $i \in I$ where $b_i$ indicates the bid giving from $i$ and $b_{-i}$ means others biding and $v_i$ denotes the true value for $i$. 
\begin{equation}
    U_i(b_i,b_{-i}) = v_i^m - c_i^m - Tax(b_i,b_{-i},c)
\end{equation}

\emph{\textbf{Theorem 1 (Truth telling).} Telling the true value is the weak dominant strategy under the proposed Clarke Tax.}
\begin{equation}
    U_i(v_i,b_{-i}) \geq U_i(b_i',b_{-i})
\end{equation}

The prove for Theorem 1 is presented in Appendix A.\\

 Intuitively, the dishonest participant will pay the tax at least equal to the extra acquirement from lying due to the Clarke tax mechanism.

\subsubsection{Responsibility of the model}

In order to encourage the model owner to be responsible to the model its proposed, a mechanism of punishment is needed in case the actual accuracy doesn't agree with the expected one after $t$ rounds of training. An practicable method is to let the rest of model owner to vote for whether to punish the winner. This approach is established based on the assumption that all participants are honest and are not in the competitive relation which make no incentive to give a malicious vote. More sophisticated approach is implement reputation mechanism which has been discussed in previous work.

\subsubsection{Tax recycle}
In Clarke Tax mechanism, a potential drawback is the waste of the tax. To leverage the tax revenue, the smart contract can offer a discount or price rising depending on the side of the tax payer for the payment in the next training round.

\subsection{Feasibility}

\subsubsection{Budget Balance}

The total revenue is the sum of payment from the owners and the tax obtained from both side of participants and the expenditure is the rewards for the client. Due to the fact that a model will not be trained unless its sum of the payment exceed the total rewards, the smart contract can therefore easily decide a payment $C_m$ which independent to the claim and budget balance.\\

\emph{\textbf{Theorem 2 (Budget Balance).} The proposed VCG-based crowdfunding mechanism is budget balance.} 

\textit{Proof} The total revenue $R$ in round from the model owners can be written as:
\begin{equation}
    R = \sum c_i^m + \sum Tax(b_i,b_{-i},c) \geq C_m
\end{equation}{}

Therefore, the budget is balanced.\\

\subsubsection{Participation constraint}
Although the participants are not aware of the actual payment before its execution, the overall social utility will be positive. That means the expectation of the utility for the participants are positive. \\

\emph{\textbf{Definition 1 (Risk Neutral).} The expected utility for risk neutral participants is equal to the utility under expected crowdfunding outcome.} 

\begin{equation}
    EU =U(E(m)) 
\end{equation}
where $m$ denotes the crowdfunding model and its connected properties(e.g. cost, tax).\\
Note that, the average utility is positive in crowdfunding scenario.\\

\begin{equation}
    Avg(U(m)) = \frac{1}{|O|}(\sum v^m_i - C_m )\geq 0, i \in O
\end{equation}
\begin{equation}
    Avg(U(m)) = \frac{1}{|B|}( C_m - \sum v^m_i )\geq 0, i \in B
\end{equation}

For the risk neutral participants, they can estimate their expectation of utility before joining the crowdfunding. Therefore, the participant will take part in the crowdfunding if it could reach the average level(e.g. common target, capability, contribution).\\
\indent
Furthermore, the determination of the final payment is a matter of the belief of fairness and all participants are aware of it (i.e. $v^i_{E(m)},c^i_{E(m)}$) before join the crowdfunding. It not only can be equally divide to everyone but also can be allocated depending on the capability or contribution which can provide incentive in FL training. Therefore, the individual who participate in the crowdfunding all owns a positive expected utility.
\begin{equation}
    v^i_{E(m)} \geq c^i_{E(m)}, i \in O
\end{equation}
\begin{equation}
    c^i_{E(m)}\geq v^i_{E(m)} , i \in B
\end{equation}

\subsection{Clients' Payment and Incentive}
Note that the reward pool for client is settle before the training process. To provide incentive for client in training, the smart contract can allocate the reward by the contribution of the client. The contribution can be quantified as the margin accuracy of the local model. Due to the fact that the accuracy owns a close connection to the data quality, the clients can assess the expectation of the reward based on their data quality in order to determine whether to join in the cooperation. Furthermore, consequently, the client have the incentive to improve the data quality and reduce their cost, which virtually rises the efficiency of the HCFL.

\section{Experimental}
\begin{figure*}
	\setlength{\abovecaptionskip}{0pt}
	\setlength{\belowcaptionskip}{0pt}
	\centering
	\begin{minipage}[t]{0.48\linewidth}
		\centering
		\includegraphics[width=2.5in]{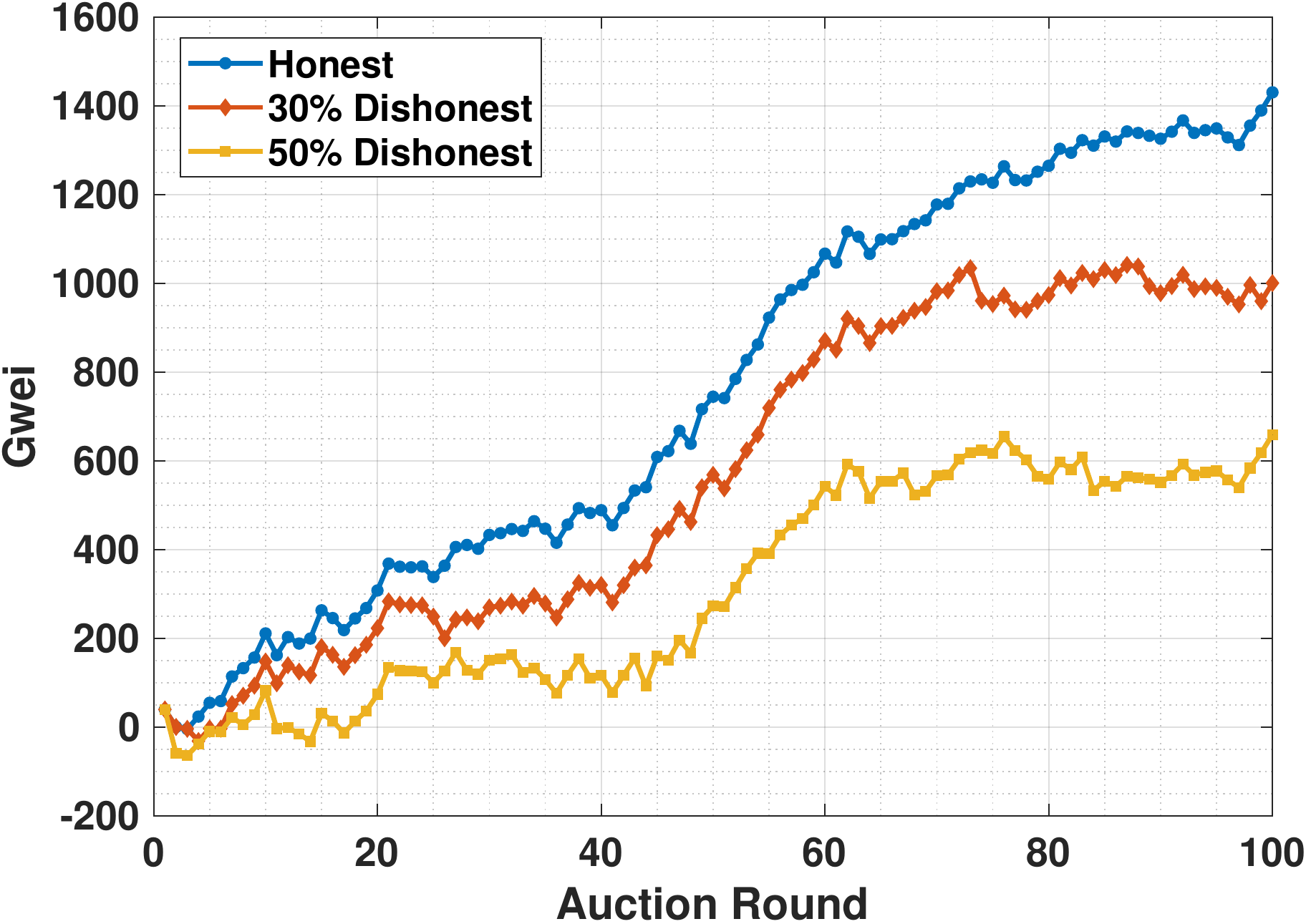}
\caption{\scriptsize{The accumulated utility comparison between truthfulness and dishonesty under VCG auction.}}
		\label{fig:GBD_conver}
	\end{minipage}%
	\begin{minipage}[t]{0.48\linewidth}
		\centering
		\includegraphics[width=2.5in]{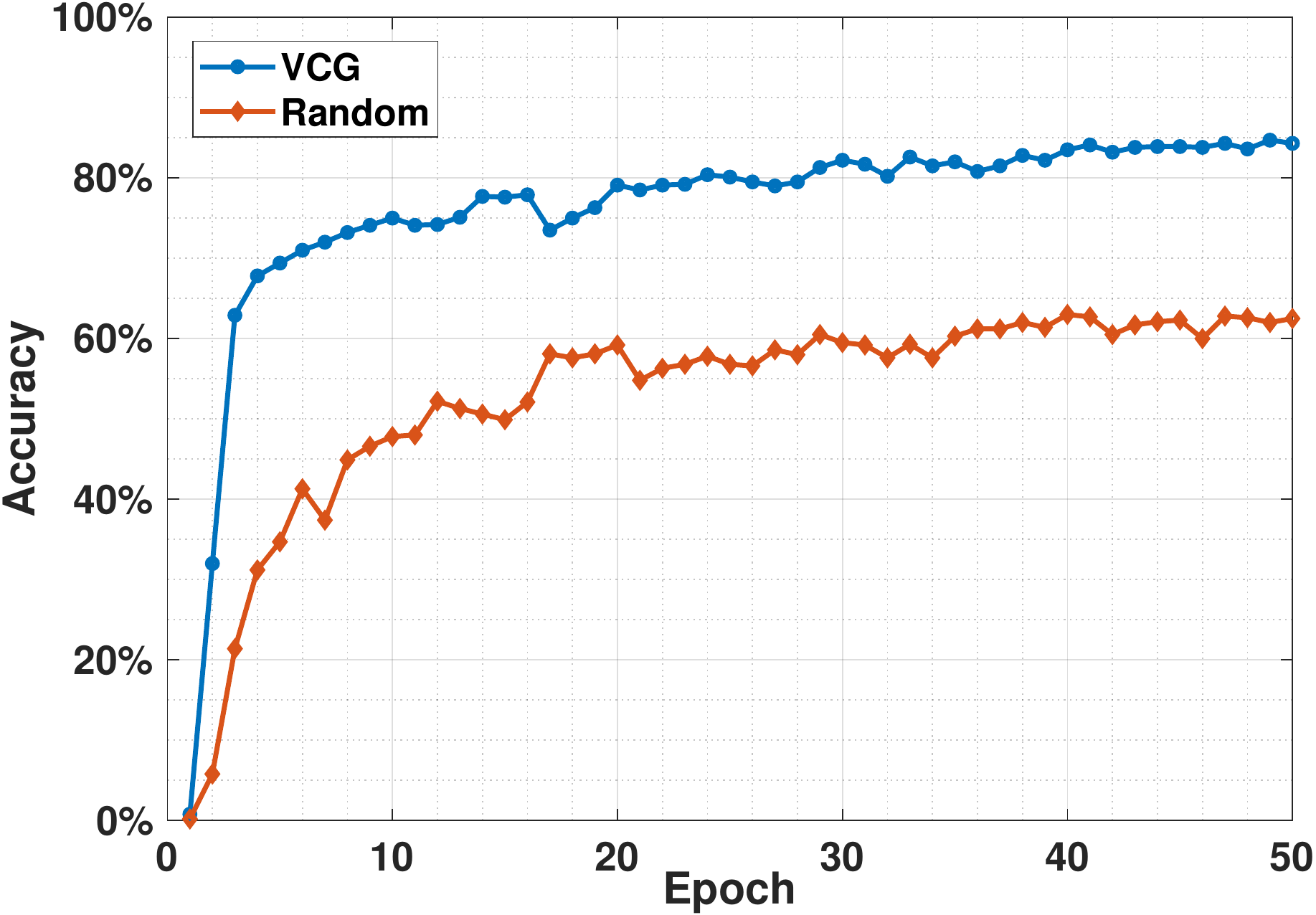}
\caption{\scriptsize{The expected accuracy for commonly interested target in Cifa10.}}
		\label{fig:incentives}
	\end{minipage}
	\begin{minipage}[t]{0.48\linewidth}
		\centering
		\includegraphics[width=2.5in]{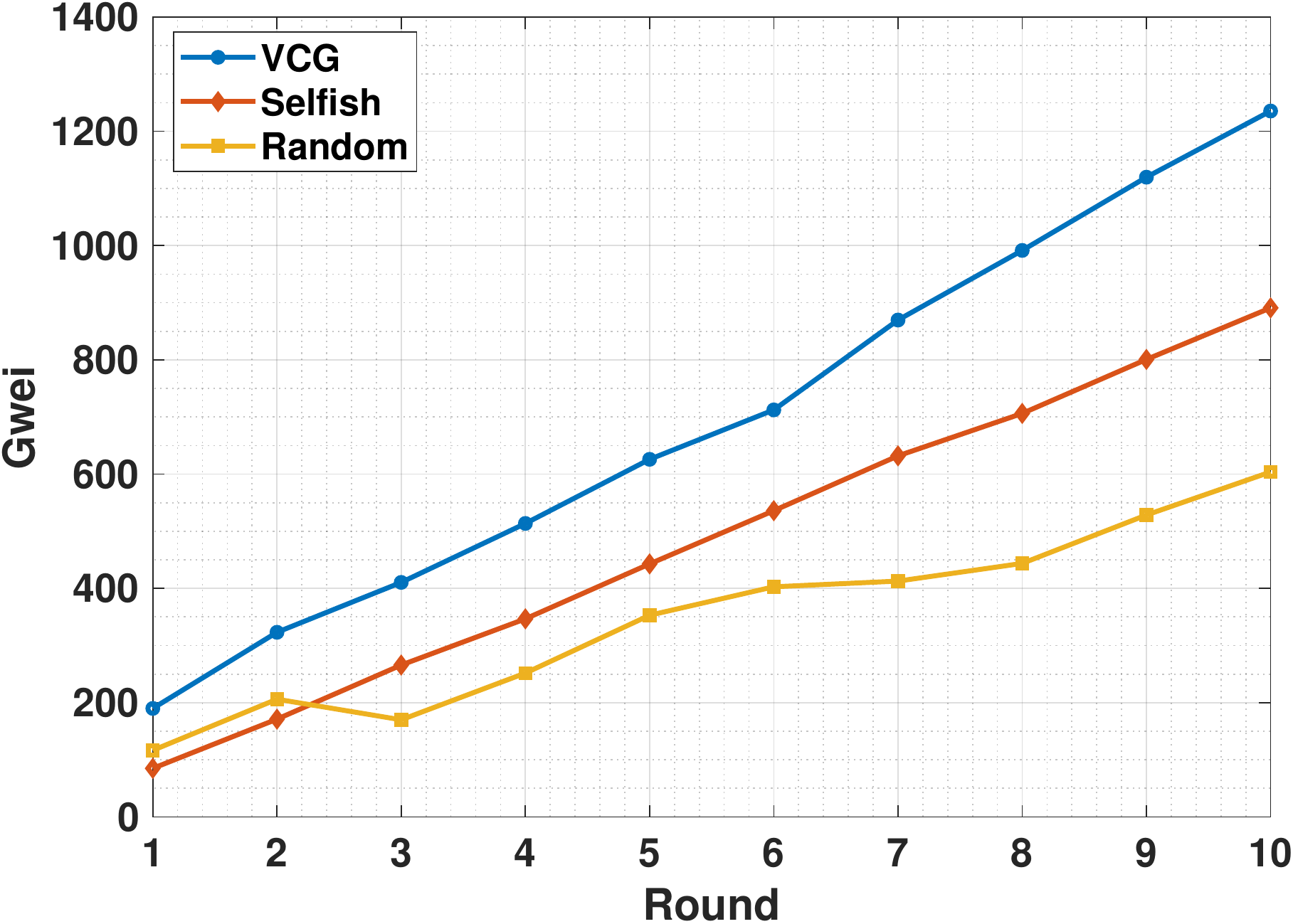}
\caption{\scriptsize{The accumulated social utility under three different strategies.}}
		\label{fig:accu_ener}
	\end{minipage}%
	\begin{minipage}[t]{0.48\linewidth}
 		\centering
		\includegraphics[width=2.5in]{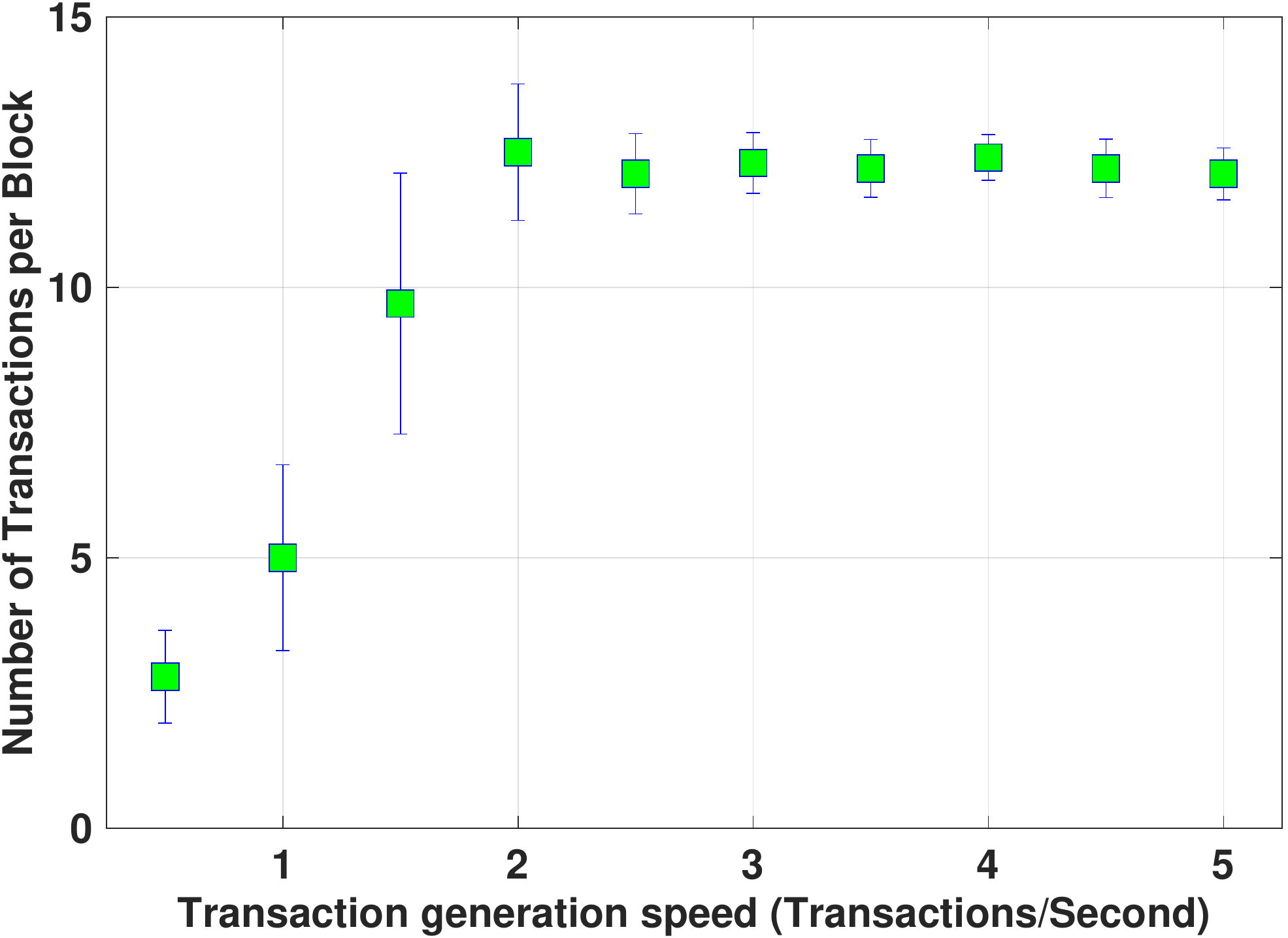}
\caption{\scriptsize{The performance of the proposed smart contract.}}
		\label{fig:smart_contract}
	\end{minipage}%
\end{figure*}





In this section,  we first implement the the HCFL framework and exam the effectiveness of the smart contract. Then, we demonstrate the proposed VCG mechanism enable truth telling to become the weak dominate strategy. Finally, we verify the model selected and trained in HCFL can achieve the maximum social utility.  \\
\indent
In the experiment, we simulate 10 model owners and 10 base stations. For simplicity, the model is fixed as Resnet50 but the target of the model is to be determined by auction. The whole system is deployed on an Intel Core CPU Xeon Gold 5222 with a clock rate 3.80 GHz with 16 cores and two threads per core. The FL model is written in Python 3.8.10 and Pytorch 1.10.1. The execution operates on the Geforce RTX 2080Ti GPU.

\subsection{Participants' Utility}

In this section, we demonstrate the truthfulness of the VCG mechanism by  comparing the different utility between truth-telling and dishonesty. The dishonesty is illustrated by the possibility of the participant claiming a fake value.  We set the mean value of the model owner to be 100 Gwei(the current unit in Ethereum) and 50 Gwei for base sation. Then, we randomly choose a group of true value for each of them as the proposed model assessments. The model auction is conducted under the rule of the proposed VCG mechanism and repeated for 100 round. The utility of  participant in each round can be calculated as the margin between the true value and the final payment. Fig.\ref{fig:GBD_conver} shows the accumulated utility for a random model owner in two strategies(i.e. honesty or dishonesty).

\indent
It is apparent that the utility from truth telling is greater than dishonesty in any stage. Furthermore, it is important to note that the utility can be negative in a single round and causing fluctuation in the plot. However, negative utility denotes that the commonly support model is not profitable for the particular participant. If the cooperation union is effective, the expectation of the utility will be positive which has discussed in Section 4.

\subsection{Cooperation Performance}

This section exams the selected model that can perform well in commonly interested target.\\
\indent
we simulate 10 model owners with an random intersection from all labels in Cifa10 to denote the common interests. Then, we randomly expand the label set as the individual valued target and, consequently, the proposed model target. Then, we implement the VCG auction and select the model that can maximize the over all utility. The base line is randomly chosen model target. To equalize the cost between the VCG group and the base line and simulate the resource-limited scenario, we assume the training cost is closely related to the scale of the data set and set the scale of the training data set to be 10000 graphs in Cifa10. The average accuracy is present in Fig.\ref{fig:incentives} after 50 global epochs with 2 local epochs in each round. \\
\indent
Shown in Fig.\ref{fig:incentives}, the accuracy on the intersection is noticeable greater than that on the base line. These indicates that proposed mechanism in this article can leverage limited resource to maximize the most valued utility, which is suitable for those participants whose targets are similar but limited with resource.

\subsection{Social Utility}
This section, we illustrate the proposed VCG mechanism can lead to the efficient social utility. we assign a target set from the labels in Cifa10 to each model owner. For each participant, each label in the target set, the value and cost is randomly selected with mean value of 10 Gwei for model owner and  5 Gwei for base station. The smart contract will determine the model under by three strategies: 1) Random selection 2)Selfish 3)VCG auction. The selfish strategy is to find the intersection among the model owners.  The social utility is quantified as the margin between the total value and the cost multiply by the average accuracy of the model after training. After 10 rounds of training, the accumulated social utility is shown in Fig.\ref{fig:accu_ener}. In the experiment, we guarantee the existence of the intersection in order to conduct the selfish strategy in each round, which may overrate the capability of the selfish strategy.\\
\indent
Note that the accumulated social utility of the VCG auction is always larger than the other two strategies. This verifies that the VCG mechanism is able to select the model that can achieve the maximum social utility from all model proposals.

\subsection{Smart Contract Performance}
This section is aimed to validate the feasibility of deploying the VCG mechanism based on smart contract considering the transaction efficiency.\\
\indent
Fig.\ref{fig:smart_contract} shows the performance of the proposed smart contract. As can be seen from Fig.\ref{fig:smart_contract}, the number of transactions contained in the block increases with the rate of task generation and subsequently reaches saturation when the transaction generation speed is around 2 transactions per second. This indicates that the proposed framework is able to operate under a acceptable transaction speed, which validates the effectiveness of the proposed scheme.

\section{Future Work}
\subsection{Model Evaluation}
In this article, we assume the participants can rationally evaluate the value and the cost of each proposed model. However, for the clients and base station, the cost for each model is complicated. On the another hand, for the model owner, how to select a desirable model that can fulfill the expectation is remain to explore.

\subsection{Incentive}
Although many works have discussed the incentive mechanism in FL, the situation in this scenario has a little difference that the total rewarding pool is pre-determined before the training. That means no matter how the eventual performance of the model is, the total payment for the clients is the same which may cause failure if a considerable amount of clients dawdle. We have discussed the incentive and punishment in the Discussion section, but the details of the particular mechanism is remained to be refined.

\subsection{Feasibility in other scenario}
The scenario in this article is limited in cooperation among the participants who owns similar target, which means it is more suitable for horizontal federated learning.
The feasibility in vertical federated learning is remained to be verified. Furthermore, crowdfunding with VCG mechanism may be potentially practicable in other scenario where individual needs but lacks resource. For instance, in mobile cloud computing, the individual device is resource-hungry but resource-constrained\cite{Chen2015}.

\section{Conclusion}
In this article, we introduce a smart contract based hierarchy federated learning architecture(HCFL) which enable capability-limited model owners to cooperate truthfully and efficiently in a crowdfunding scenario. A smart contract is deployed in  HCFL to enforce the procedure and preserve the equal statue between the model owners and the base stations. Leveraging the  VCG mechanism, the smart contract of HCFL can obtain  the honest claim of assessment from both model owner and the base station and, consequently, is able to determine the model that could bring the greatest social utility. We conduct a thorough discussion of the properties of HCFL and implement several simulations to demonstrate the effectiveness of the mechanism which indicates that the HCFL can benefit the participants who are cooperative in  this crowdfunding scenario.  

\section{Appendix}
\subsection{Demonstration for Theorem 1}
Here, we offer the demonstration of the validity of the Clarke Tax mechanism(i.e. the truth telling is at least weak dominant strategy). The model owner side will the sample of this demonstration and that of device side is similar. We start from the true telling situation and show that lying will lead to worse-off. 

\subsection*{Case 1: $\sum b_i^{m} \geq c_{m}, (\sum\limits_{I / i} b_i^m- \sum\limits_{I / i} c_j^m) \geq  (\sum\limits_{I / i} b_j^{m'}- \sum\limits_{I / i} c_j^{m'})$  and $(v_i^m + \sum\limits_{I / i} b_j^{m}-c_{m}) \geq    (v_i^{m'}+ \sum\limits_{I / i} b_j^{m'} - c_{m'})$}

Here, the participant $i$ is not the pilot for it does not affect the final decision. Therefore, the payment is $c_i^{m}$.
The utility for $i$ is $u_i = v_i^{m} - c_i^{m}$. If $i$ try to vary its bid to change the outcome(i.e. when $c_i^m < v_i^m$ giving a bid $b_i^{m'} > b_i^m = v_i^m$  it will remain to be not the pilot and pay is still $v_i^{m} - c_i^{m}$. However, if $c_i^m > v_i^m $ and it bids a price $ b_i^{m'} < b_i^m = v_i^m$ and eventually change the decision, it becomes the pilot and pays  $(|\sum\limits_{I / i} b_i^{m} - \sum\limits_{I / i} C_i^{m}|-|\sum\limits_{I / i} b_i^{m'}- \sum\limits_{I / i} C_i^{m'}|)$

we then need prove:
\begin{equation}
    v_i^{m} - c_i^{m} \geq  v_i^{m'} - c_i^{m'}-(( \sum\limits_{I / i} b_j^{m} - \sum\limits_{I / i} c_j^{m})-(\sum\limits_{I / i} b_j^{m'}- \sum\limits_{I / i} c_j^{m'})), 
\end{equation}{}

\begin{equation}
    \begin{array}{c}
         v_i^m \geq v_i^{m'}+ C_m - C_{m'}- \sum\limits_{I / i} b_j^{m} + \sum\limits_{I / i} b_j^{m'} \\
         (v_i^m + \sum\limits_{I / i} b_j^{m}-C_{m}) \geq    (v_i^{m'}+ \sum\limits_{I / i} b_j^{m'} - C_{m'})
    \end{array}{}
\end{equation}{}

Therefore, for $i$, the weak dominant strategy is to be honest(e.g. $v_i^m = b_i^m$).

\subsubsection*{Case 2: $\sum b_i^{m'} \geq c_{m'}, (\sum\limits_{I / i} b_j^m- \sum\limits_{I / i} c_j^m) \leq  (\sum\limits_{I / i} b_j^{m'}- \sum\limits_{I / i} C_j^{m'})$  and $(v_i^m + \sum\limits_{I / i} b_j^{m}-C_{m}) \geq    (v_i^{m'}+ \sum\limits_{I / i} b_j^{m'} - C_{m'})$}

In this situation, $i$ is the pilot if it is honest. The utility for it is 
\begin{equation}
    u_i = v_i^m - c_i^m + (\sum\limits_{I / i} b_j^{m} - \sum\limits_{I / i} c_j^{m})- (\sum\limits_{I / i} b_j^{m'}- \sum\limits_{I / i} c_j^{m'}) 
\end{equation}{}

Then, by rearranging the order, we have

\begin{equation}
    u_i= v_i^m - C_m + \sum\limits_{I / i} b_j^{m} - \sum\limits_{I / i} b_i^{m'} + C_{m'} \geq v_i^{m'} - c_i^{m'} 
\end{equation}
 
If it bids a higher price than $v_i^m$, the utility remains the same. And is it bids a lower price and cause the variation of the final decision, the current model will fail from the selection. According to Eq.(3), this is not a rational strategy.

Therefore, the weak dominant strategy for $i$ is still to be honest.

\subsubsection*{Case 3: $\sum b_i^{m} \geq c_{m}, (\sum\limits_{I / i} b_j^m- \sum\limits_{I / i} c_j^m) \geq  (\sum\limits_{I / i} b_j^{m'}- \sum\limits_{I / i} c_j^{m'})$  and $(v_i^m + \sum\limits_{I / i} b_j^{m}-C_{m}) \leq    (v_i^{m'}+ \sum\limits_{I / i} b_j^{m'} - C_{m'})$ }

In  this situation the participant $i$ is also the pilot. However, the model fails in the selection which leads to pure tax for $i$. Its utility is:
\begin{equation}
    u_i = v_i^{m'}-c_i^{m'}-(\sum\limits_{I / i} b_j^{m} - \sum\limits_{I / i} c_j^{m})+ (\sum\limits_{I / i} b_j^{m'}- \sum\limits_{I / i} c_j^{m'})
\end{equation}{}

If it continually lower its bid, the utility won't change. On the contrary, if it rises its bid and change the outcome(e.g. $m$ wins in the selection). The utility for it is:
\begin{equation}
    u_i' = v_i^m - c_i^m
\end{equation}{}
\begin{equation}
    u_i - u_i' = C_m + \sum\limits_{I / i} b_j^{m'} -\sum\limits_{I / i} b_j^{m}- C_{m'} +v_i^{m'} -v_i^m \geq 0
\end{equation}{}

Therefore, the weak dominant strategy for $i$ is still to be honest.

\subsubsection*{Case 4: $\sum b_i^{m'} \geq c_{m'}, (\sum\limits_{I / i} b_j^m- \sum\limits_{I / i} c_j^m) \leq  (\sum\limits_{I / i} b_j^{m'}- \sum\limits_{I / i} c_j^{m'})$  and $(v_i^m + \sum\limits_{I / i} b_j^{m}-C_{m}) \leq    (v_i^{m'}+ \sum\limits_{I / i} b_j^{m'} - C_{m'})$ }

Similar to Case 1, the utility is
\begin{equation}
    u_i = v_i^{m'} - c_i^{m'}
\end{equation}{}

If $i$ rises its price and become the pilot, its utility is
\begin{equation}
    u_i' = v_i^{m} - c_i^{m} + (\sum\limits_{I / i} b_j^{m} - \sum\limits_{I / i} c_j^{m})- (\sum\limits_{I / i} b_j^{m'}- \sum\limits_{I / i} c_j^{m'})
\end{equation}{}
\begin{equation}
    u_i - u_i' = v_i^{m'} - v_i^{m} + C_m - C_{m'} - \sum\limits_{I / i} b_j^{m}+\sum\limits_{I / i} b_j^{m'} \geq 0
\end{equation}{}

Therefore, the weak dominant strategy for $i$ is still to be honest.

\subsubsection*{Case 5: $\max\limits_{M}(\sum b_i^{m}- c_{m}) < 0$ }
Apparently, the utility for $i$ is $u_i = 0$.
if it rises its price(e.g. $b_i^m > v_i^m$) and become the pilot
\begin{equation}
    u_i' = v_i^m - c_i^m - \sum\limits_{I / i}c_i^m + \sum\limits_{I / i}b_i^m = v_i^m + \sum\limits_{I / i}b_i^m - C_m < 0
\end{equation}{}

Therefore, the weak dominant strategy for $i$ is still to be honest.\\
\indent
In conclusion, this Clarke Tax mechanism can guarantee the truthful biding.
\bibliographystyle{IEEEtran}
\bibliography{IEEEabrv,vcg_review}

\end{document}